\documentclass[prb,floatfix,showpacs,superscriptaddress,twocolumn]{revtex4}
\usepackage{amsmath}
\usepackage{amssymb}
\usepackage{amsfonts}
\usepackage[pdftex]{graphicx,hyperref}
\usepackage{dcolumn}
\usepackage{bm}
\usepackage{chngcntr}
\usepackage{changes}

\DeclareGraphicsExtensions{.pdf,.png,.jpg,.eps}

\begin{document}

\title{Collective modes of trapped spinor Bose condensates}
\author{C. Trallero-Giner}
\affiliation{Department of Theoretical Physics, Havana University, Havana 10400, Cuba}
\affiliation{Departamento de F\'{\i}sica, Universidade Federal de S\~{a}o Carlos,
13.565-905 S\~{a}o Carlos, Brazil}
\author{Dar\'{\i}o G. Santiago-P\'{e}rez}
\affiliation{Universidad de Sancti Spiritus \textquotedblleft Jos\'{e} Mart\'{\i} P\'{e}
rez", Ave. de los M\'{a}rtires 360, CP 62100, Sancti Spiritus, Cuba}
\affiliation{CLAF - Centro Latino-Americano de F\'{\i}sica, Avenida Venceslau Braz, 71,
Fundos, 22290-140, Rio de Janeiro, RJ, Brasil}
\author{V. Romero-Rochin}
\affiliation{Instituto de F\'{\i}sica, Universidad Nacional Aut\'{o}noma de M\'{e}xico,
Apartado Postal 20-364, 01000 M\'{e}xico, Distrito Federal, M\'{e}xico}
\author{G. E. Marques}
\affiliation{Departamento de F\'{\i}sica, Universidade Federal de S\~{a}o Carlos,
13.565-905 S\~{a}o Carlos, Brazil}
\date{\today }

\begin{abstract}
We study the richer structures of quasi-one-dimensional Bogoliubov-de Genes
collective excitations of $F=1$ spinor Bose-Einstein condensate in a
harmonic trap potential loaded in an optical lattice. Employing a
perturbative method we report general analytical expressions for the
confined collective polar and ferromagnetic Goldstone modes. In both cases
the excited eigenfrequencies are given as function of the 1D effective
coupling constants, trap frequency and optical lattice parameters. It is
shown that the main contribution of the optical lattice laser intensity is
to shift the confined phonon frequencies. Moreover, for high intensities,
the excitation spectrum becomes independent of the self-interaction
parameters. We reveal some features of the evolution for the Goldstone modes
as well as the condensate densities from the ferromagnetic to the polar
phases.
\end{abstract}

\pacs{03.75.Be, 03.75.Lm, 05.45.Yv}
\maketitle

\section{\label{Introduction}Introduction}

Since the pioneering works of Ho~\cite{PhysRevLett.81.742} and Ohmi and
Machida,~\cite{doi:10.1143/JPSJ.67.1822} a lot of effort have been devoted
to study $F=1$ spinor Bose-Einstein condensates (BECs). Due to the internal
degrees of freedom, the condensate presents a vectorial character and the
wavefunction is described by three components in the hyperfine state with
magnetic quantum number $m=$-1,0,1. Accordingly and depending of
spin-exchange interaction values,~\cite
{PhysRevA.64.053602,PhysRevLett.88.093201} two phases are predicted:
ferromagnetic ($^{87}$Rb)~\cite%
{Chang2005,PhysRevLett.83.2498,PhysRevLett.87.010404} and polar or
antiferromagnetic ($^{23}$Na).~\cite{PhysRevLett.80.2027,PhysRevLett.82.2228}
 The experimental realization of spinor BECs, typically produced in
optically trapped dilute gases, allows to study several interesting problems
as magnetism, quantum phenomena not observed in single-component condensate,~
\cite{Stenger1998,PhysRevLett.82.2228} spin dynamics,~\cite
{PhysRevLett.92.140403} the miscibility of the spinor components~\cite
{Stenger1998} as well as spin domains in an external magnetic field, and the
nature of the ground state spinor condensates (see the recently overview in
Ref.~\onlinecite{Kawaguchi2012253} and references there in).

An analysis of the collectives modes is an important step for a compressive
study of the dynamics for both polar and ferromagnetic phases. Problems as
quantized vortices,~\cite{PhysRevLett.83.2498} superfluidity, spin-domains~
\cite{Stenger1998} or damping processes, require an exhaustive knowledged of
the dynamic process (see Refs.~
\onlinecite{PhysRevA.92.042502,PhysRevA.88.063638,PhysRevA.70.023611,
PhysRevA.77.033610,1367-2630-11-12-123012}). Information of the collective
excitations on the atom-atom self-interaction terms and on the applied
external potential are fundamental bricks to build the dynamics of the
phenomena above mentioned. The nature and evaluation of the Goldstone modes,
so far, has been tackled assuming a spatial homogeneous Bose-Einstein
condensate.~\cite%
{PhysRevLett.81.742,doi:10.1143/JPSJ.67.1822,PhysRevLett.110.235302,Kawaguchi2012253}
Thus, due to the concomitant translational symmetry the wavevector $\mathbf{k
}$\textbf{\ }is a good quantum number with Bogoliubov typical excitation $
\omega =\omega (\mathbf{k}).$ However, this approach is not longer valid
when the condensate is loaded into a confined spatial trap potential, in
particular, the collective excitations must show a discrete set of modes or
confined states.

In the present contribution we focus on the behavior of the Goldstone modes
for $F=1$ spinor BECs confined in a cigar-shape geometry. The experimental
realizations of two- and one-dimensional condensates in diluted ultracold
atoms employing optical and magnetic traps are very well established
technics.~\cite{PhysRevLett.87.130402,PhysRevLett.87.160405,Science293} In
general, however, the three dimensional nonlinear Gross-Pitaevskii equation
(GPE) cannot be decoupled into transversal and longitudinal motions.
Nevertheless, in presence of highly anisotropic trap potential, we can
handled the problem as being tightly confined in a plane and with an
independent 1D motion.~\cite{Petrov} Ref.~\onlinecite{PhysRevLett.81.742}
shows that the three-dimensional ground state $\Psi _{m}$ of the alkali
atoms of the condensate in the hyperfine state $\left\vert m\right\rangle $ $
(m=-1,0,1)$ is ruled by

\begin{multline}
i\hslash \frac{\partial \Psi _{m}(x,\mathbf{r},t)}{\partial t}=\left[ \frac{
p^{2}}{2M}+U_{3D}+\right. \\
\left. \overline{c_{0}}\Psi _{a}^{\dagger }\Psi _{a}\right] \Psi _{m}(x,
\mathbf{r},t)+\overline{c_{2}}\Psi _{a}^{\dag }\mathbf{F}_{ab}\Psi _{b}\cdot
\left[ \mathbf{F}\Psi (x,\mathbf{r},t)\right] _{m}\text{ },  \label{eq:GP3D}
\end{multline}
with $M$ being the mass of the atom, $\mathbf{F}$ the total hyperfine spin
operator$,$ $\overline{c_{0}}=(g_{0}+2g_{2})/3$, $\overline{c_{2}}
=(g_{2}-g_{0})/3$, $g_{j}$ ($j=0,2$) the atom-atom self-interaction terms
related to the s-wave scattering length, $a_{F}$ in the total spin $F$
channel and, additionally, depending on the total number of particles $N$.
The order parameter can be written explicitly as $\Psi _{a}(x,\mathbf{r},t)=$
$\left\vert \Psi (x,\mathbf{r},t)\right\vert \zeta _{a}$, where $\zeta _{a}$
is the normalized spinor state and $\int_{{\mathbb{R}}}\left\vert \Psi
_{a}(x,\mathbf{r},t)\right\vert ^{2}\,dxd\mathbf{r}=1.$ Above, we have
written $(x,\mathbf{r})$ as the 3D spatial vector position, with $x$ being
the longitudinal coordinate and $\mathbf{r}$ the 2D transverse vector.

In Eq.~(\ref{eq:GP3D}) we consider a condensate confined in anisotropic
harmonic trapping potential and loaded into an optical lattice as given by
the external potential

\begin{equation}
U_{3D}(x,\mathbf{r})=\frac{1}{2}M\left( \omega _{0}^{2}x^{2}+\omega _{
\mathbf{r}}^{2}\mathbf{r}^{2}\right) -V_{L}\cos ^{2}(\frac{2\pi }{d}x)\text{
},  \label{3Dpotentail}
\end{equation}
where $\omega _{0}$ and $\omega _{\mathbf{r}}$ are the longitudinal and
transversal harmonic oscillator frequencies, respectively, $V_{L}$ is
proportional to the laser intensity and $d$ the laser wavelength. Assuming
now that the longitudinal motion is adiabatic with respect to the transverse
one, and considering a highly anisotropic harmonic trapping, with $\omega
_{0}$ $\ll $ $\omega _{\mathbf{r}}$, the 3D order parameter can be
factorized into~\cite{PhysRevA.74.023607} $\Psi _{m}(x,\mathbf{r},t)=\phi
_{m}(x;t)\chi (x,\mathbf{r})$. We note two important consequences of these
assumptions, one, the transverse motion is independent of the hyperfine
state $\left\vert m\right\rangle $ and, second, all the time evolution
occurs along the 1D longitudinal coordinate $x$. It then follows from Eqs.~(
\ref{eq:GP3D}) and (\ref{3Dpotentail}) that the motion in the plane is given
by the equation~\cite{PhysRevLett.89.110401,refId0}

\begin{multline}
\left[ \frac{p_{\mathbf{r}}^{2}}{2M}+\frac{M}{2}\omega _{\mathbf{r}}^{2}
\mathbf{r}^{2}+\left( \overline{c_{0}}\left\vert \phi _{a}\right\vert ^{2}+
\frac{1}{\phi _{m}}\overline{c_{2}}\phi _{a}^{\dag }\mathbf{F}_{ab}\phi
_{b}\cdot \left( \mathbf{F}\phi \right) _{m}\right) \times \right.  \\
\left. \left\vert \chi (x,\mathbf{r})\right\vert ^{2}\right] \chi =\mu _{
\mathbf{r}}\left[ \phi _{m}\right] \chi \text{ },  \label{transm}
\end{multline}
where the transverse chemical potential, $\mu _{\mathbf{r}}\left[ \phi _{m}
\right] $, is a functional of the 1D order parameter $\phi _{m}(x;t)$. For
the longitudinal dynamic part we have

\begin{multline}
i\hslash \frac{\partial \phi _{m}(x;t)}{\partial t}=\left[ \frac{p_{x}^{2}}{
2M}+\frac{M}{2}\omega _{0}^{2}x^{2}-\right. \\
\left. V_{L}\cos ^{2}(\frac{2\pi }{d}x)+\mu _{\mathbf{r}}\left[ \phi _{m}
\right] \right] \phi _{m}(x;t)\text{ }.  \label{longmo}
\end{multline}
In a first approach the function $\chi (x,r)$ can be described by the ground
state of a 2D harmonic oscillator with frequency $\omega _{\mathbf{r}}$.
Thus, expanding $\mu _{\mathbf{r}}$ as a Taylor series of the wavefunction $
\phi _{m}$ and following the same trend as giving in Refs.~{
\onlinecite{refId0} and \onlinecite{carretero} we have }

\begin{equation}
\mu _{\mathbf{r}}\left[ \phi _{m}\right] \phi _{m}\approx \hslash \omega _{
\mathbf{r}}\phi _{m}+C_{m}\phi _{m}+.....\text{ },  \label{2dAppr}
\end{equation}
with $C_{m}\phi _{m}=c_{0}\left\vert \phi _{m}\right\vert ^{2}\phi
_{m}+c_{2}\phi _{a}^{\dag }\mathbf{F}_{ab}\phi _{b}\cdot \left( \mathbf{F}
\phi \right) _{m}$, $c_{0}=M\omega _{\mathbf{r}}\overline{c_{0}}/(3\hslash
\pi )$ and $c_{2}=M\omega _{\mathbf{r}}\overline{c_{2}}/(3\hslash \pi )$
becoming the effective self-interaction constants for the 1D cigar-shape
spinor BEC. Hence, Eq.~(\ref{longmo}) is reduced to

\begin{multline}
i\hslash \frac{\partial \phi _{m}(x;t)}{\partial t}=\left[ \frac{p_{x}^{2}}{
2M}+\frac{M}{2}\omega _{0}^{2}x^{2}-V_{L}\cos ^{2}(\frac{2\pi }{d}x)+\right.
\\
\left. c_{0}\left\vert \phi _{m}\right\vert ^{2}\right] \phi
_{m}(x;t)+c_{2}\phi _{a}^{\dag }\mathbf{F}_{ab}\phi _{b}\cdot \left( \mathbf{
F}\phi \right) _{m}\text{ }.  \label{1D}
\end{multline}

Starting now from the 1D spinor GPE, Eq.~(\ref{1D}), the main contribution
of this work is the description of the collective longitudinal modes,
providing explicit expressions for the corresponding excited wavefunctions, $
\varphi _{m,k}(x;t)$, and their eigenfrequencies, $\hslash \omega _{m}^{(k)}$
($k=1,2,\dots $). In Sec.~\ref{Excitedstates} we consider the quasi-1D
generalized Bogoliubov-deGennes equations (B-dGEs), which allows for the
analysis of the polar and ferromagnetic phases and the interplay between the
non-linear terms, on one hand, and the harmonic trapping, and the optical
lattice external potentials on the other one. As we stated above, the system
is consider as spatially inhomogeneous and in consequence we are in presence
of confined Goldstone modes or confined phonon like spectrum. Section~\ref
{WaveFunction} is devoted to the excitation amplitudes for both phases
considered, and our conclusions are listed in Sec.~\ref{Conclusion}. The
main elements for the description of the eigenfrequencies and eigenfunctions
are summarized in the Appendixes~\ref{appendix A} and~\ref{appendix B},
respectively.

\section{\label{Excitedstates}Excited states}

Using Bogoliubov approximation at very low temperature,~\cite{Pitaevskii}
the collective excitation states of the generalized 1D GPE~(\ref{1D}) can be
represented by the wavefunction having the form

\begin{multline}
\varphi _{m,k}(x;t)=\text{exp}(-i\mu t/\hbar )\left[ \phi _{0}(x)\right. \\
\left. +\varphi _{m}^{(k)}\text{exp}(-i\omega _{m}^{(k)}t)+\varphi _{\mp
m}^{\dag (k)}\text{exp}(i\omega _{m}^{(k)}t)\right] \text{,}
\label{BogoliubovAprop}
\end{multline}
where $\varphi _{m}^{(k)}$ represents a small fluctuation from the
stationary solutions $\phi _{0}(x)\text{exp}(-i\mu t/\hbar )$ with the sign (
$\mp )$ for the polar ($P$) and the ferromagnetic ($Fe)$ phases,
respectively and $\mu $ the 1D chemical potential. Following Eq.~(\ref
{BogoliubovAprop}), the wavefunctions of the collective modes for the
condensate satisfy the generalized 1D B-dGEs~\cite
{Pitaevskii,PhysRevLett.81.742}

\begin{equation}
i\hbar \frac{\partial }{\partial t}\left(
\begin{array}{c}
\varphi _{m}^{(k)} \\
-\varphi _{\mp m}^{\dag (k)}
\end{array}
\right) =\left(
\begin{array}{cc}
L_{0} & \overline{\lambda _{m}^{(J)}}|\phi _{0}|^{2} \\
\overline{\lambda _{m}^{(J)}}|\phi _{0}|^{2} & L_{0}
\end{array}
\right) \left(
\begin{array}{c}
\varphi _{m}^{(k)} \\
\varphi \varphi _{\mp m}^{\dag (k)}
\end{array}
\right) \text{ },  \label{BdG}
\end{equation}
where $J$ =($P$, $Fe$) and the operator $L_{0}$ is defined as
\begin{equation}
L_{0}=\frac{1}{2M}p_{x}^{2}+U-\mu +\lambda _{m}^{(J)}|\phi _{0}|^{2}\text{ }.
\end{equation}
In Eq.~\ref{BdG}) and (\ref{L0}), the coupling parameters are given by,

\begin{equation}
\lambda _{m}^{(P)}=\left\{
\begin{array}{cc}
2c_{0}; & m=0 \\
g_{2}; & m=\pm 1
\end{array}
\right. \text{, \ \ }\overline{\lambda _{m}^{(P)}}=\left\{
\begin{array}{cc}
c_{0}; & m=0 \\
c_{2}; & m=\pm 1
\end{array}
\right. \text{ }.  \label{lamnda}
\end{equation}
and

\begin{eqnarray}
\lambda _{m}^{(Fe)} &=&\left\{
\begin{array}{cc}
g_{2}; & m=0 \\
g_{2}+2\left\vert c_{2}\right\vert ; & m=-1 \\
2g_{2}; & m=1
\end{array}
\right. \text{ ,}  \label{lamndaF} \\
\text{\ }\overline{\lambda _{m}^{(Fe)}} &=&\left\{
\begin{array}{cc}
0; & m=-1,0 \\
g_{2}; & m=1
\end{array}
\right. \text{ }.
\end{eqnarray}

\begin{figure}[tbh]
\begin{center}
\includegraphics[width=\columnwidth]{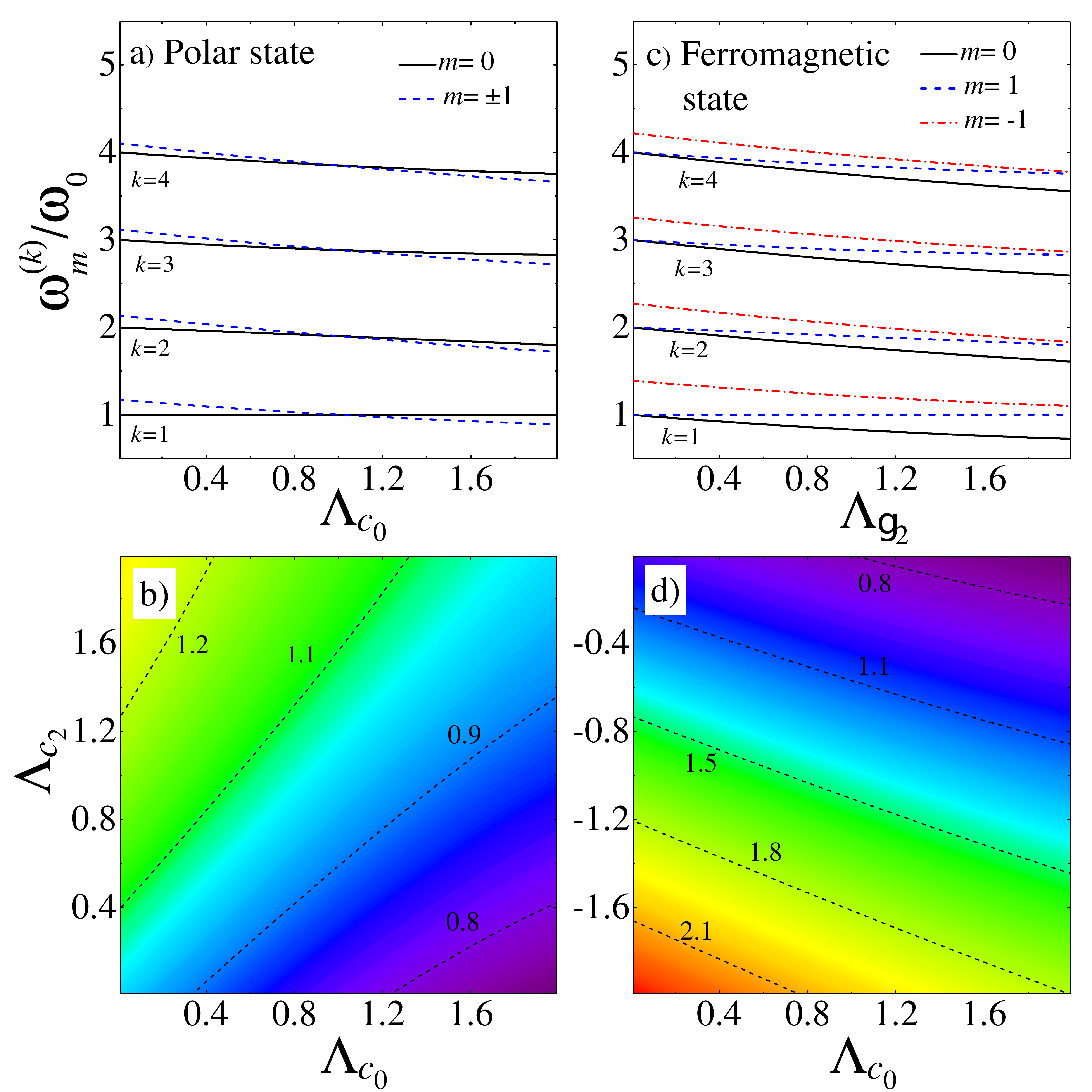}
\end{center}
\caption{(Color online) Upper panel: Collective excitation frequencies $
\protect\omega _{m}^{(k)}$, in units of the trap frequency potential $
\protect\omega _{0}$, of the first four modes. a) $\protect\omega
_{P,m}^{(k)}$ as function of $\Lambda _{c_{0}}$; b) $\protect\omega
_{Fe,m}^{(k)}$ as function of $\Lambda _{g_{2}}$. Lower panel: Contour maps
as function of 1D self-interaction constants $\Lambda _{c_{0}}$ and $\Lambda
_{c_{2}}$: c) $\protect\omega _{P,\pm 1}^{(1)}$; d) $\protect\omega
_{Fe,1}^{(1)}$. In the calculation $V_{L}=0$.}
\label{PolarF}
\end{figure}
The order parameter $\phi _{0}$ is described by the GPE

\begin{equation}
\left[ \frac{1}{2M}p_{x}^{2}+U(x)+g_{1}|\phi _{0}|^{2}\right] \phi _{0}=\mu
\phi _{0}\text{ },  \label{GPE}
\end{equation}
with $U(x)=M\omega _{0}^{2}x^{2}/2-V_{L}\cos ^{2}(2\pi x/d)$,

\begin{equation*}
g_{1}=\left\{
\begin{array}{ccc}
c_{0} & \rightarrow & \text{polar state} \\
g_{2} & \rightarrow & \text{ferromagnetic state}
\end{array}
\right. \text{ .}
\end{equation*}
We search for the solution of Eq.~(\ref{GPE}) with the boundary conditions $
\phi _{0}\rightarrow 0$ at $x\rightarrow \pm \infty $ and normalized to $
\int |\phi _{0}(x)|^{2}dx=1.$ Analytical expressions for the order
parameter, $\phi _{0},$ and the chemical potential, $\mu ,$ solution of Eq.~(
\ref{GPE}) for a given self-interaction constant $g_{1}$, are reported in
Ref.~\onlinecite{PhysRevA.79.063621}.

\subsection{Polar phase}

\subsubsection{Homogeneous System.}

First, we consider a negligible intensity for the external trap potential
and without an optical lattice, in which the system can be considered as
homogeneous. In this situation, the density $n_{0}$ is constant within all
space and the system presents spatial symmetry invariance. Hence, the linear
momentum $p_{x}$ is a good quantum number. Solving the system of Eqs.~(\ref
{BdG}), for the phonon-like Bogoliubov excitation spectrum in the
low-momentum regime, we have that $\hslash \omega _{P,0}=\sqrt{\varepsilon
_{p_{x}}(\varepsilon _{p_{x\text{ }}}+2c_{0}n_{0})}$ and $\hslash \omega
_{P,\pm 1}=\sqrt{\varepsilon _{p_{x}}(\varepsilon _{p_{x}}+2c_{2}n_{0})}$
with $\varepsilon _{p_{x}}=p_{x}^{2}/2M$.~\cite
{Bogolyubov,PhysRevLett.79.4056,Pitaevskii}

\subsubsection{Confined Phonons.}

If we now tackled the problem with the external 1D trap potential $M\omega
_{0}^{2}x^{2}/2\neq 0$, the spatial symmetry invariance is broken and we are
facing to a discreet set of confined phonon-like modes with frequencies $
\omega _{P,m}^{(k)}$ $(k=1,2,...)$. Equations~(\ref{GPE}) and (\ref{BdG})
form an independent 3$\times $3 system of equations for $m=0$ and $\pm 1$.
By considering both, the nonlinear terms $\overline{\lambda _{m}^{(P)}}|\phi
_{0}|^{2}$ and the optical lattice potential in the B-dGEs~(\ref{BdG}) as a
perturbation with respect to the harmonic trap, we are able to get the
collective phonon mode frequencies, $\omega _{P,m}^{(k)}$, for each
hyperfine state $\left\vert m\right\rangle $. A description of the employed
perturbative algorithm is given elsewhere.~\cite
{EPJDTrallero2012,PhysRevA.92.042502} According to the values of the
interaction constant $\lambda _{m}^{(P)}$, the inherent symmetry of the
system~(\ref{BdG}) shows that the states with $m=\pm 1$ are degenerate. The
corresponding analytical results for the eigenfrequencies $\omega
_{P,m}^{(k)}$ are displayed in the Appendix~\ref{appendix A}. In the upper
panel of Fig.~\ref{PolarF}a) we show $\omega _{P,m}^{(k)}$ in units of $
\omega _{0}$ for the first 4 modes and $m=0$, $\pm 1$ as a function of the
dimensionless interaction parameter $\Lambda _{c_{0}}=c_{0}/(l_{0}\hbar
\omega _{0})$ for the repulsive case $\Lambda _{c_{0}}>0$. Here $l_{0}=\sqrt{
\hslash /(M\omega _{0})}$. We observe that $\omega _{P,0}^{(1)}=\omega _{0}$
is constant independent of the self-interaction constants,~\cite{Pitaevskii}
while the other modes $\omega _{P,m}^{(k)}$ decreasing as $\Lambda _{c_{0}}$
increases. It is interesting to note that for $\Lambda _{c_{0}}>\Lambda
_{c_{2}}=c_{2}/(l_{0}\hbar \omega _{0})$ the first excited state correspond
to $\omega _{P,\pm 1}^{(1)}$ and in general $\omega _{P,\pm 1}^{(k)}>\omega
_{P,0}^{(k)},$ $\forall $ $k$ (see Appendix~\ref{appendix A} Eqs.~(\ref{P0})
and (\ref{P1})). In Fig.~\ref{PolarF}b) the evolution of the collective
excitation $\omega _{P,\pm 1}^{(1)}$ is shown in terms of the interactions $
\Lambda _{c_{0}}$ and $\Lambda _{c_{2}}$. For given values of the parameter $
\Lambda _{c_{0}}$ we observe that the frequency increases monotonically as $
\Lambda _{c_{2}}$ increases. In these calculations we fixed the intensity of
the optical lattice as $V_{L}=0$.

\subsection{Ferromagnetic phase}

\subsubsection{Homogeneous System}

This phase emerges when $c_{2}<0$ and we have three set of non-degenerate
states $\varphi _{m}^{(k)}(x)$ for $m=-1,0,1$. As in the Polar case, the
energies of the excited states are obtained directly from the Eqs~(\ref{GPE}
) and (\ref{BdG}) and can be cast as $\hslash \omega _{Fe,-1}=\varepsilon
_{p_{x}}+2c_{2}n_{0}$, $\hslash \omega _{Fe,0}=\varepsilon _{p_{x}}$, and $
\hslash \omega _{Fe,1}=\sqrt{\varepsilon _{p_{x}}(\varepsilon
_{p_{x}}+2g_{2}n_{0})}$. Here, the only phonon-like Bogoliubov spectrum
corresponds to the state $\varphi _{1}(x)$ with frequency $\omega _{Fe,1}$.

\subsubsection{Confined Phonons}

\begin{figure}[tbh]
\begin{center}
\includegraphics[width=\columnwidth]{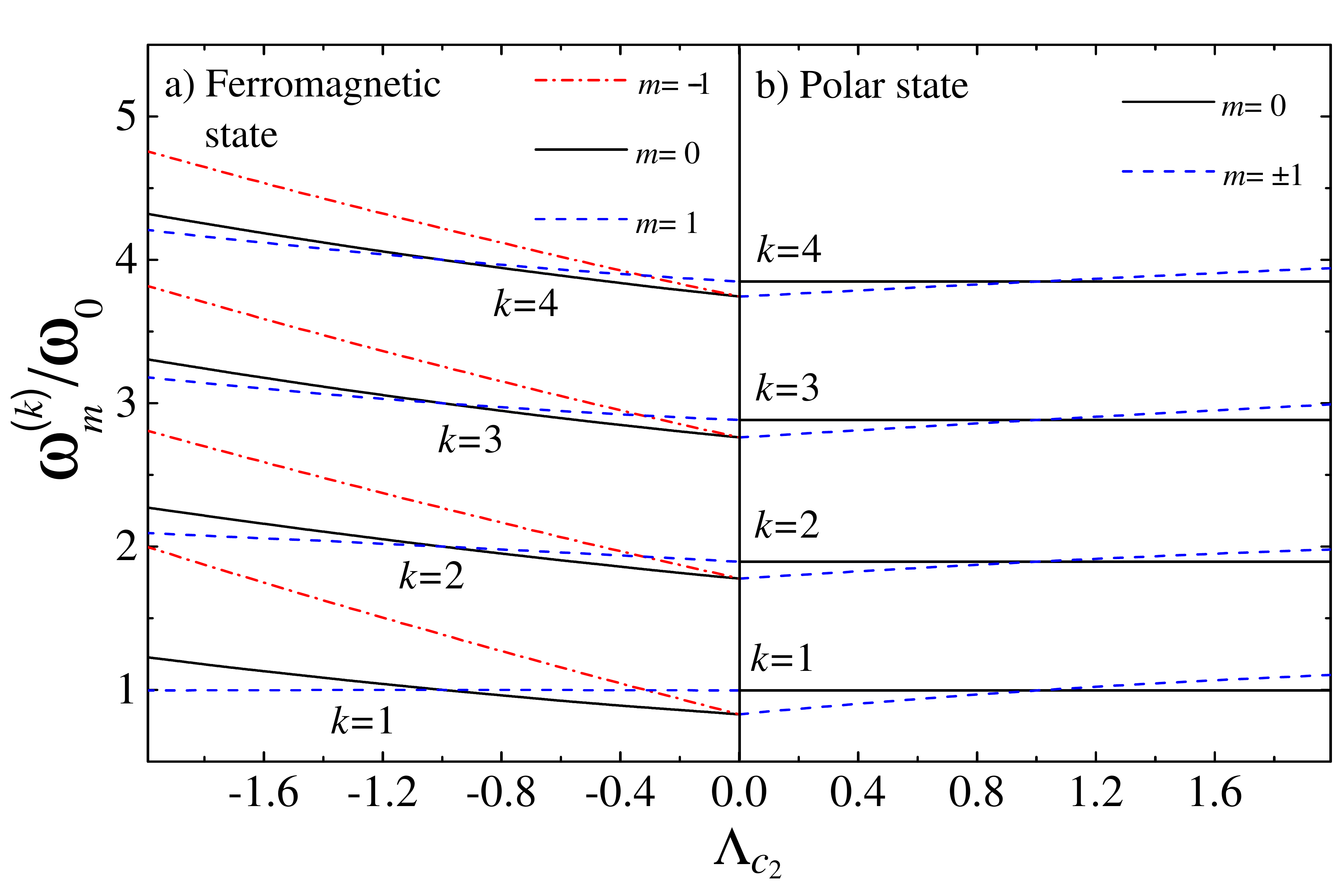}
\end{center}
\caption{(Color online) Evolution of the frequencies modes ($k=1,2,3$, and 4
) from the ferromagnetic phase to the polar phase as a function $\Lambda
_{c_{2}}$. Left (right) panel $\Lambda _{g_{2}}=1$ ($\Lambda _{c_{0}}=1$).}
\label{FerroMasPolarFrequency}
\end{figure}

In the present case, the system~(\ref{BdG}) is decoupled into two
independent equations for $\varphi _{Fe,0}^{(k)}$ and $\varphi
_{Fe,-1}^{(k)} $, and $3x3$ system of equations for the state with $m=1$.
Following the same procedure mentioned above for the Polar phase, in
Appendix~\ref{appendix A} we report the analytical solutions for the three
independent excited frequencies $\omega _{Fe,m}^{(k)}$, $m=-1,0,1$, $\forall
$ $k$. Figures~\ref{PolarF}b) and d) are devoted to the collective
excitations for the ferromagnetic phase. In the upper panel of the figure we
observe the three independent set ($m=-1,0,1$) of confined frequencies ($
k=1,2,3,4$) as a function of $\Lambda _{g_{2}}=g_{2}/(l_{0}\hbar \omega _{0})
$. All frequencies decrease as $\Lambda _{g_{2}}$ increases, while the state
$(m=1,k=1)$ is independent of the interaction constants.Notice that the
states with $m=-1,0$ do not fulfill typical properties of B-dGE solutions,
for instance, their first excited state is independent of the interaction.
This appears to be in correspondence with the fact that in the homogeneous
case, their dispersion relations are not linear in the low-momentum regime..
In~\ref{PolarF}d) the characteristic contour map for the reduced confined
phonon frequency $\omega _{Fe,-1}^{(1)}/\omega _{0}$ is represented as a
function of $\Lambda _{c_{0}}$ and $\Lambda _{c_{2}}$. For a given value of $
\Lambda _{c_{0}}$ the frequency $\omega _{Fe,-1}^{(1)}$ decreases as\ $
\Lambda _{c_{2}}\rightarrow 0$ in correspondence with the result shown in
Fig.~\ref{FerroMasPolarFrequency}, as discussed below.

\begin{figure}[tbh]
\begin{center}
\includegraphics[width=\columnwidth]{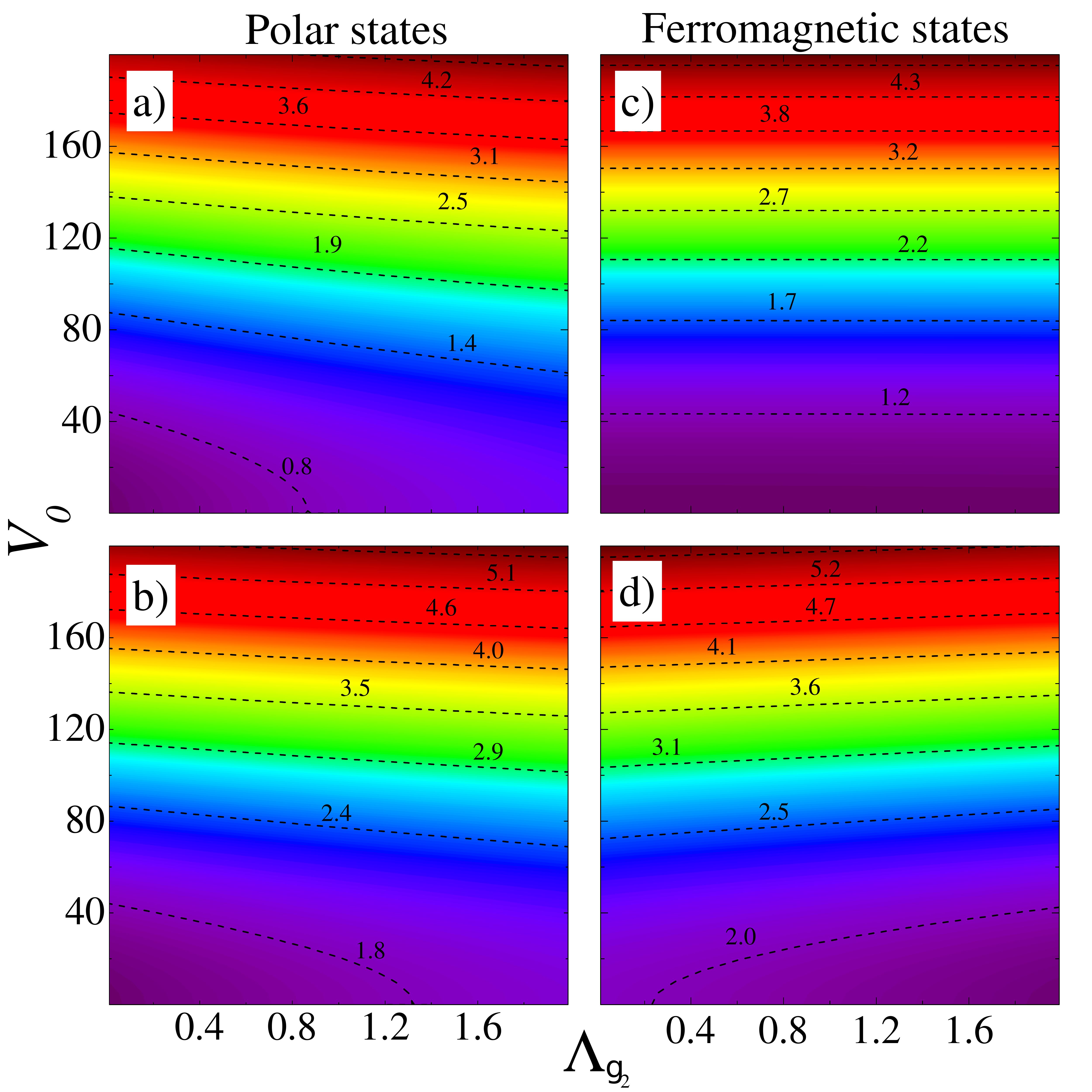}
\end{center}
\caption{(Color online) Contour plot of the excited frequencies $\protect
\omega _{m}^{(k)}$ as function of dimensionless parameters $\Lambda _{g_{2}}$
and $V_{0}=V_{L}/\hslash \protect\omega _{0}$ for the first two modes $k=1,2$
. Left panel: Polar state for $\protect\omega _{\pm 1}^{(k)}$. Right panel:
Ferromagnetic state and $\protect\omega _{1}^{(k)}$.}
\label{Curvasnivelvo}
\end{figure}

The transition from the ferromagnetic to the polar phase is represented in
Fig.~\ref{FerroMasPolarFrequency} for the modes with frequencies $\omega
_{m}^{(k)}$ ($k=1,2,3,4$) as function of $\Lambda _{c_{2}}$. In the panel a)
of the figure for $\Lambda _{c_{2}}<0$, the three set of independent modes
with $m=-1,0,-1$ are very well resolved$.$ They show different behavior as $
\Lambda _{c_{2}}$ decreases with the stronger slope for the phonon modes $
\omega _{-1}^{(k)}$. For $\Lambda _{c_{2}}=0$, the states $m=-1,0$ become
degenerate, while for $\Lambda _{c_{2}}$ positive, the values of $\omega
_{P,m}^{(k)}$ are closer to $k\omega _{0}$, (see panel a)), i.e. the
influence of $\Lambda _{c_{2}}$ is negligible and we have that these three
states are quasi-degeneracy.

An important issue is the influence of the optical lattice on the collective
excitations. Figure~\ref{Curvasnivelvo} shows contour plot of the
frequencies $\omega _{m}^{(k)}$ for the first two states ($k=1,2$) as a
function of the dimensionless laser intensity\ $V_{L}$ and the parameter $
\Lambda _{c_{2}}$ for the polar state, $m=\pm 1$, and the ferromagnetic one,
$m=1$. The main contribution of $V_{L}$ is to shift the confined phonon
frequency. For larger values of $V_{L}$, the frequency is almost independent
of $\Lambda _{c_{2}}$, while the mayor modification of $\omega _{m}^{(k)}$
occurs for lower values of laser intensity, $V_{L}\sim 40\hbar \omega _{0}$.
These facts are explained by Eqs.~(\ref{m=+/-1}) and (\ref{m=1}) that take
into account the interplay between the self-interaction constant $\Lambda
_{c_{2}}$ and the presence of the optical lattice.

\section{\label{WaveFunction}Excitation amplitudes}

The wavefunction of the excited states for the polar and ferromagnetic phase
are displayed in the Appendix~\ref{appendix B}. The calculation of $\varphi
_{m}^{(k)}(x;t)=\varphi _{m}^{(k)}$exp$(-i\omega _{m}^{(k)}t)$ is obtained
in first order of perturbation for the self-interaction constants $\Lambda
_{c_{0}}$, $\Lambda _{c_{2}},$ $\Lambda _{g_{2}}$ and dimensionless laser
intensity $V_{0}=V_{L}/\hslash \omega _{0}$. As it is states in the Appendix~
\ref{appendix B}, the space of solutions $\varphi _{m}^{(k)}(x)$ is composed
of two independent Hilbert subspaces $\mathcal{H}_{\mathcal{I}}$ and $
\mathcal{H}_{\mathcal{II}}$ for odd ($k=1,3,...)$ and even ($k=2,4,$...)
wavefunctions with respect to the inversion symmetry $x\rightarrow -x.$

The condensate density perturbation for a given phonon frequeny $\omega
_{m}^{(k)}$ can be cast as

\begin{equation}
\delta \varphi _{m}^{(k)}=\frac{|\varphi _{m,k}(x;t)|^{2}-|\phi _{0}(x)|^{2}
}{\sqrt{N}}\text{ }.  \label{density}
\end{equation}%
Thus, employing the results of the Appendix \ref{appendix B} for the
wavefunction of the excited states of the polar state with $m=0$, we obtain
an analytical representation for the function $\delta \varphi _{0}^{(k)}$,
given by

\begin{multline}
\delta \varphi _{0}^{(k)}=2\text{cos}(\omega _{0}^{(k)}t)\left[ \phi _{0}(x)
\overline{\phi _{k}(x)}-\frac{\text{exp}(-x^{2})}{\sqrt{\pi ^{\frac{1}{2}}}}
\times \right.  \label{densityPerturbation} \\
\left( \Lambda _{c_{0}}\frac{f_{k,0}}{k\sqrt{\pi ^{\frac{1}{2}}}}+\left\{
\sum_{m\neq k}\Lambda _{c_{0}}\left( \frac{2}{k-m}-\frac{1}{k+m}\right)
f_{k,m}\right. \right. \\
\left. \left. \left. -V_{0}\frac{g_{k,m}}{2(k-m)}\right\} \overline{\phi
_{k}(x)}\right) \right] \text{ }.
\end{multline}

\begin{figure}[tbh]
\begin{center}
\includegraphics[width=\columnwidth]{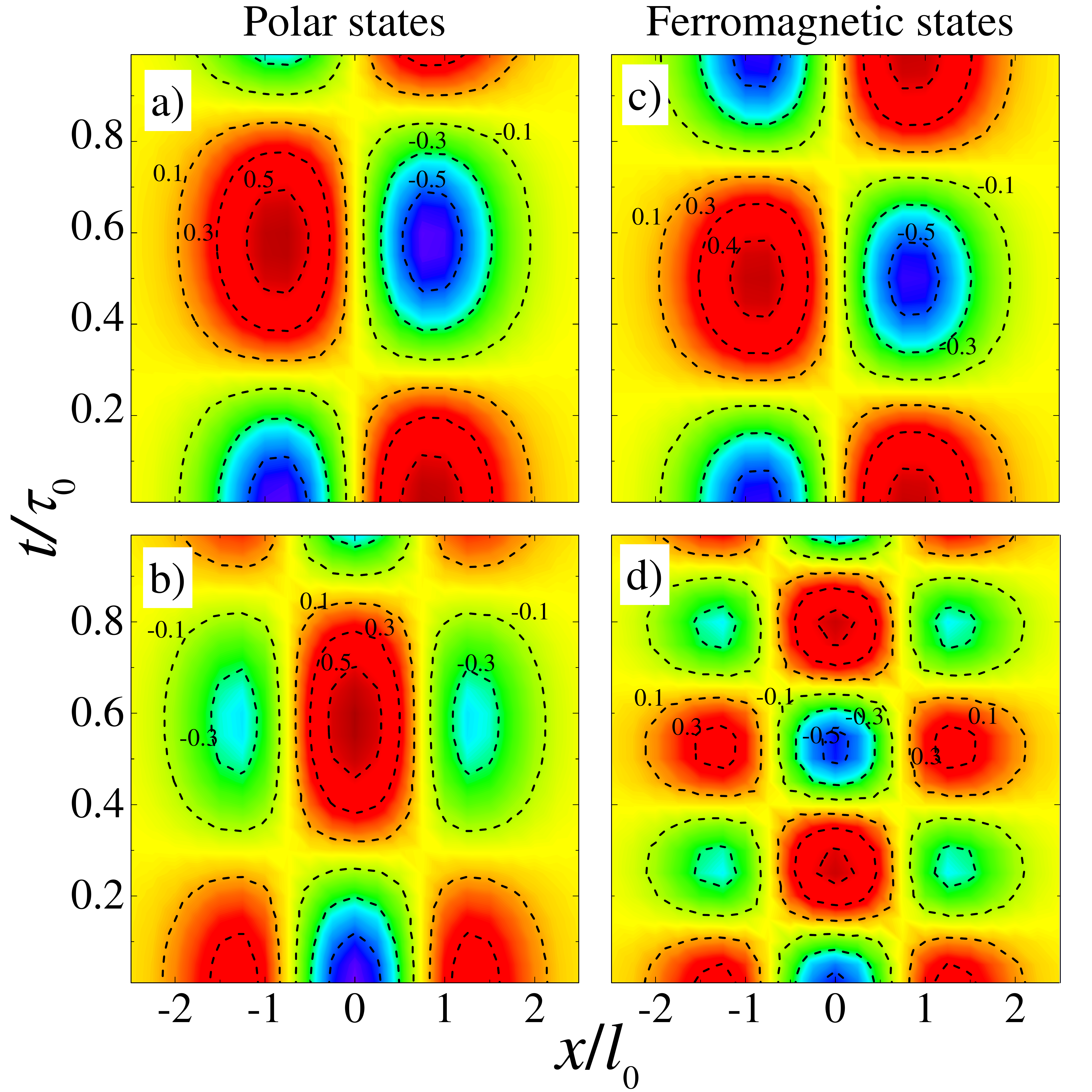}
\end{center}
\caption{Contour map of the 1D condensate density perturbation, $\protect
\delta \protect\varphi _{1}^{(k)}(x,t)$, for the first two excited state $
(k=1,2)$. Left panel: Polar modes for $\Lambda _{c_{2}}=0.5$ and $\Lambda
_{g_{2}}=2$. Right panel: Ferromagnetic modes for $\Lambda _{g_{2}}=1$. In
the calculation $\protect\tau _{0}=2\protect\pi /\protect\omega _{0}$ and $
V_{L}=0$ }
\label{FluctuacionUno}
\end{figure}
Similar results can be obtained for the Bogoliubov-type excitation
amplitudes listed in the Appendix~\ref{appendix B}. In Fig.~\ref
{FluctuacionUno} it is shown a contour plot of the condensate density
perturbation $\delta \varphi _{1}^{(k)}(x;t)$ for the polar and
ferromagnetic phases. Here, we consider the first two excited states, the
first one with $k=1$ belongs to the Hilbert subspaces $\mathcal{H}_{\mathcal{
I}}$, while for $k=2$ to $\mathcal{H}_{\mathcal{II}}.$ The antisymmetric and
symmetric character of $\delta \varphi _{1}^{(k)}$ for both phases, are
clearly seen in the figure. In general, the evolution from one phase to
another as a function of the parameter $\Lambda _{c_{2}}$ does not change
the parity of a the density perturbation $\delta \varphi _{m}^{(k)}(x;t)$.

\section{\label{Conclusion}Conclusion}

In conclusion, we have solved the multicomponent order parameter of the
coupled Bogoliubov-de Genes equations, Eq.~(\ref{BdG}), for the one
dimensional cigar-shaped Bose-Einstein condensates with $F=1$ spin degrees
of freedom.\ We have presented useful analytical expressions for the
confined phonon frequencies and wavefunctions of the excited states for the
ferromagnetic and antiferromagnetic phases. The examen of the Goldstone
modes shows that the phonon energies, in both polar and ferromagnetic
phases, are proportional to the longitudinal harmonic trap frequency. We
conclude that the phonon modes are weakly dependent on the interaction
constants for the antiferromagnetic states, while a more pronounced
structure is reached in the case of BEC loaded in the ferromagnetic phase
(see Fig.~(\ref{PolarF}) and (\ref{FerroMasPolarFrequency})). Also, we found
the existence of a set of the self-interaction constant values for which the
lower frequency lies below of the harmonic oscillator frequency $\omega _{0}$
. The modes for the polar $m=0$ and ferromagnetic $\ m=-1$ states coincide
with the oscillation of the center of mass and are independent of the
atom-atom interactions.~\cite{Pitaevskii} In contrast to results obtained in
the framework of Thomas-Fermi approximation, where the density of excited
polar modes are interaction independent (see Refs.~
\onlinecite{PhysRevLett.81.742} and~\onlinecite{PhysRevLett.77.2360}), we
have found here that the condensate densities, $\delta \varphi _{m}^{(k)},$
show a clear structure and depend on the\ $g_{0}$ and $g_{2}$ atom-atom
self-interaction terms.

\appendix

\section{\label{appendix A}Exited frequencies}

Introducing the dimensionless interaction self-interaction constant $\Lambda
_{C}=C/(l_{0}\hbar \omega _{0})$ ($C=c_{0},c_{2}$, and $g_{2}$), $l_{0}=
\sqrt{\hslash /(M\omega _{0})}$, $V_{0}=V_{L}/\hslash \omega _{0}$ and $
\alpha =2\pi l_{0}/d$, the eigenfrequencies, $\omega _{m}^{(k)}$, of Eq.~(
\ref{BdG}) are obtained in the framework of a perturbative regime where the
non-linear terms $\Lambda _{C}\left\vert \phi _{m}\right\vert ^{2}$ and the
periodic potential $V_{0}\cos ^{2}(\alpha x/l_{0})$ are considered as a
perturbation with respect to the trap potential. We defined the auxiliary
function

\begin{multline}
\frac{\omega ^{(k)}(z1,z2,z3)}{\omega _{0}}=k-\frac{z_{1}}{\sqrt{2\pi }}+
\frac{\Gamma (k+\frac{1}{2})}{\sqrt{2}\pi k!}z_{2}-  \label{Genefreque} \\
-\frac{V_{0}}{2}\text{exp}(-\alpha ^{2})[L_{k}(2\alpha ^{2})-1]-\frac{
z_{1}V_{0}}{\sqrt{2\pi }}\text{exp}(-\alpha ^{2})\left[ Ei\text{(}\frac{
\alpha ^{2}}{2})\right. \\
\left. -\text{ln(}\frac{\alpha ^{2}}{2})-\mathcal{C}\right] -\frac{z_{2}V_{0}
}{\sqrt{2}\pi }\text{exp}(-\alpha ^{2})\delta _{k}(\alpha )+\frac{V_{0}^{2}}{
4}\text{exp}(-2\alpha ^{2})\times \\
\left[ Chi\text{(}2\alpha ^{2})-ln\text{(}2\alpha ^{2})-\mathcal{C}+\rho
_{k}(\alpha )\right] +0.033106z_{1}^{2}+ \\
\left( \gamma _{k}^{(1)}z_{3}^{2}+\frac{\gamma _{k}^{(2)}}{4}
z_{2}^{2}\right) \frac{1}{2\pi ^{2}}\text{ },
\end{multline}%
with $\Gamma (z)$ being the Gamma function, $L_{k}(z)$ the Laguerre
polynomials, $\mathcal{C}$ the Euler's constant $Ei(z)$ and $Chi(z)$ the
exponential and cosine hyperbolic integrals, respectively, and \ $k=1,2,....$
. Functions $\delta _{k}(\alpha )$ and $\rho _{k}(\alpha )$ are reported in
Ref.~(\onlinecite{PhysRevA.92.042502}) and the values of $\gamma _{k}^{(1)}$
and $\gamma _{k}^{(2)}$ for $k=1,2,...,6$ are listed in Table I.

\begin{center}
\begin{table}[tbh]
\caption{ Values of the constants $\protect\gamma_{k}^{(1)}$ and $\protect
\gamma_{k}^{(2)}$. They satisfice the equation $\protect\gamma_{k}=\protect
\gamma_{k}^{(1)}+\protect\gamma_{k}^{(2)}$ and $\protect\gamma_{k}$ given in
Ref.~\onlinecite{EPJDTrallero2012}.}
\begin{tabular}{ccccccc}
\hline\hline
$k$ & 1 & 2 & 3 & 4 & 5 & 6 \\ \hline\hline
$\gamma_{k}^{(1)}$ & -0.284 & -0.620 & 0.142 & 0.015 & 0.093 & 0.050 \\
$\gamma_{k}^{(2)}$ & -0.486 & -0.165 & -0.162 & -0.095 & -0.079 & -0.058 \\
\hline\hline
\end{tabular}
\end{table}
\end{center}

\paragraph{Polar modes.}

Using the definition (\ref{Genefreque}), it is possible to show that the
Polar phonon modes with $m=0$ are given by
\begin{equation}
\omega _{P,0}^{(k)}=\omega ^{(k)}(\Lambda _{c_{0}},2\Lambda _{c_{0}},\Lambda
_{c_{0}})\text{ .}  \label{P0}
\end{equation}
On the other hand for the phonon frequencies $\omega _{P,\pm 1}^{(k)}$ we
have
\begin{equation}
\omega _{P\pm 1}^{(k)}=\omega ^{(k)}(\Lambda _{c_{0}},\Lambda
_{g_{2}},\Lambda _{c_{2}})\text{ }.  \label{P1}
\end{equation}

\paragraph{Ferromagnetic modes.}

The confined phonon frequencies for the $m=0$ ferromagnetic states can be
cast as
\begin{equation*}
\omega _{Fe,0}^{(k)}=\omega ^{(k)}(\Lambda _{g_{2}},\Lambda _{g_{2}},0)\text{
}.
\end{equation*}
The modes with $m=1$ have the eigenfrequencies
\begin{equation*}
\omega _{Fe,1}^{(k)}=\omega ^{(k)}(\Lambda _{g_{2}},2\Lambda
_{g_{2}},\Lambda _{g_{2}})\text{ },
\end{equation*}
and for $m=-1$ we obtain
\begin{equation*}
\omega _{Fe,-1}^{(k)}=\omega ^{(k)}(\Lambda _{g_{2}},\Lambda
_{g_{2}}+2\left\vert \Lambda _{c_{2}}\right\vert ,0)\text{ }.
\end{equation*}

\section{\label{appendix B} Wavefunction of the excited states}

In first order of perturbation $\Lambda _{g_{2}}$ and $V_{0}$ we obtain the
eigensolutions $\varphi _{m}^{(k)}$. Firstly, we introduce the auxiliary
function

\begin{multline}
\digamma _{m}^{(k)}(y,t;z_{1},z_{2})=\overline{\phi _{k}(y)}\exp (-i\omega
_{m}^{(k)}t)+  \label{F} \\
\sum_{p\neq k}\left[ \frac{4z_{1}f_{k,p}(y)-V_{0}g_{k,p}(y)}{2(k-p)}\right.
\\
\left. \overline{\phi _{p}(z)}\exp (-i\omega _{m}^{(k)}t)\right] - \\
\sum_{p=0}^{\infty }\frac{z_{2}f_{k,p}(y)}{k+p}\overline{\phi _{p}(y)}\exp
(i\omega _{m}^{(k)}t)\text{ ,}
\end{multline}
where $y=x/l_{0}$, $\overline{\phi _{p}(y)}$ are the 1D harmonic oscillator
wavefunctions and the functions $f_{k,p}(y),$ $g_{k,p}(y)$ are given
elsewhere.~\cite{PhysRevA.79.063621} In Eq.~(\ref{F})\ for a given state $
\left\vert m\right\rangle $, the matrix elements $f_{k,p}(y)$ and $
g_{k,p}(y) $ must fulfill the parity condition $k+p=$ even number. Thus, if $
k$ is odd $\digamma _{m}^{(k)}$ is antisymmetric, while for $k$ even the
function (\ref{F}) is symmetric. In consequence, the density perturbation $
\delta \varphi _{m}^{(k)}$ is restricted by the symmetry property of the $
\digamma _{m}^{(k)}$ function.

For the polar state the excited wavefunction with $m=0$ is reduced to
\begin{equation}
\varphi _{P;k,m=0}=\exp (-i\mu t/\hslash )\left[ \phi _{0}(y)+\digamma
_{0}^{(k)}(y,t;\Lambda _{c_{0}},\Lambda _{c_{0}})\right] \text{ ,}
\end{equation}
while for the case of $m\pm 1$ we obtain
\begin{multline}
\varphi _{P;k,m=\pm 1}=\exp (-i\mu t/\hslash )\left[ \phi _{0}(y)+\right.
\label{m=+/-1} \\
\left. \digamma _{\pm 1}^{(k)}(y,t;\Lambda _{g_{2}}/2,\Lambda _{c_{2}})
\right] \text{ .}
\end{multline}
For the ferromagnetic phase, the excited states are described by
\begin{multline}
\varphi _{Fe;k,m=-1}=\exp (-i\mu t/\hslash )\left[ \phi _{0}(y)+\right. \\
\left. \digamma _{-1}^{(k)}(y,t;\Lambda _{g_{2}}+2|\Lambda _{c_{2}}|,0)
\right] \text{ ,}
\end{multline}

\begin{multline}
\varphi _{Fe;k,m=0}=\exp (-i\mu t/\hslash )\left[ \phi _{0}(y)+\right. \\
\left. \digamma _{0}^{(k)}(y,t;\Lambda _{g_{2}}/2,0)\right] \text{ ,}
\end{multline}%
\begin{multline}
\varphi _{Fe;k,m=1}=\exp (-i\mu t/\hslash )\left[ \phi _{0}(y)+\right.
\label{m=1} \\
\left. \digamma _{1}^{(k)}(y,t;\Lambda _{g_{2}},\Lambda _{g_{2}})\right]
\text{ .}
\end{multline}

\acknowledgments D. S-P, C. T-G and G. E. M acknowledge support from the
Brazilian Agencies CNPq and FAPESP. C.T.-G. is grateful to the Instituto de F
\'{\i}sica, Universidad Nacional Aut\'{o}noma de M\'{e}xico, for its
hospitality. D. S-P. acknowledges support from Centro Latinoamericano de F
\'{\i}sica. C.T.-G is grateful to M.-C. Chung for useful discussions.

\end{document}